\documentclass{nature}

\usepackage{color}
\usepackage{graphicx}
\usepackage{verbatim}
\usepackage{xcolor}
\usepackage{amsmath}
\usepackage{upgreek}
\usepackage{hyperref}
\usepackage[normalem]{ulem}
\usepackage{verbatim}
\usepackage[T1]{fontenc}
\usepackage[latin2]{inputenc} 
\usepackage{gensymb}

\makeatletter

\makeatletter
\let\saved@includegraphics\includegraphics
\AtBeginDocument{\let\includegraphics\saved@includegraphics}
\renewenvironment*{figure}{\@float{figure}}{\end@float}
\makeatother

\title{A Spin Hall Ising Machine}

\author{Afshin Houshang$^{1,2,\ast}$, Mohammad Zahedinejad$^{1,2,\ast}$,  Shreyas Muralidhar$^1$, Jakub Ch\k{e}ci\'nski$^{1,3}$, Ahmad A. Awad$^{1,2}$, and Johan \AA kerman$^{1,2,\dagger}$}

\begin{document}

\maketitle

\begin{affiliations}
 \item Physics Department, University of Gothenburg, 412 96 Gothenburg, Sweden
 \item NanOsc AB, Electrum 229, 164 40 Kista, Sweden
 \item AGH University of Science and Technology, Department of Electronics, Al. Mickiewicza 30, 30-059 Krak\'ow, Poland
 
 $^\ast$These authors contributed equally to this work. 
 
 $^\dagger$E-mail: johan.akerman@physics.gu.se
\end{affiliations}

\begin{abstract}
Ising Machines (IMs) are physical systems designed to find solutions to combinatorial optimization (CO) problems mapped onto the IM via the coupling strengths of its binary spins. Using the intrinsic dynamics and different annealing schemes, the IM relaxes over time to its lowest energy state, which is the solution to the CO problem. IMs have been implemented in quantum, optical, and electronic hardware. One promising approach uses interacting non-linear oscillators whose phases have been binarized through injection locking at twice their natural frequency. Here we demonstrate such Oscillator IMs using nano-constriction spin Hall nano-oscillator (SHNO) arrays. We show how the SHNO arrays can be readily phase binarized and how the resulting microwave power corresponds to well-defined global phase states. To distinguish between degenerate states we use phase-resolved Brillouin Light Scattering (BLS) microscopy to directly observe the individual phase of each nano-constriction.
\end{abstract}

\section*{Introduction}

Conventional computers based on Von-Neumann architecture are unable to efficiently address a certain class of problems known as Combinatorial Optimization (CO) problems\cite{lawlere76}. These are by no means rare and manifest themselves in some critically important areas such as business operations, manufacturing and research, IC circuit design, protein folding and DNA sequencing, discovery of new medicines, and efficient big-data clustering, to name a few. In parallel, Moore's law continues to slow down and approach its limits making it even more vital to rethink current computation schemes and explore alternative paradigms. One important avenue in that regard is the concept of natural computing (NC) where a specific problem is mapped onto the physics of a system and lets the system converge to a stable ground state, which is the solution to the given problem. Within the realm of NC many proposal has been put forward among which quantum computers (QCs), Ising Machines, and even combinations of both, are currently the most studied examples. 

An Ising Machine (IM) is any hardware, whose node interactions can be described by an Ising Hamiltonian, and is tasked with finding its ground state, which represent the solution to a specific CO problem defined by the connection strengths between nodes. Guided by the individual interactions between all nodes, all evaluated continuously through the inherent parallelism of the system, the IM wanders through its multivariate energy landscape, typically helped by different types of annealing schemes to avoid local minima, and find its global minimum in a factorially faster time than a serial computation would. 
Given the importance of CO problems and the factorial efficiency of IMs in solving these kind of problems, there have been multiple recent hardware implementations such as quantum 
annealers\cite{Farhi472, Boixo2014}, CMOS annealers\cite{Tsukamoto2017}, nano-magnet network arrays\cite{Sutton2017}, electronic oscillators\cite{parihar2017scirep,chou2019scirep}, and laser networks\cite{Roques2019,Yamamoto2017, hamerly2019sciadv}. Launched in 2011, D-Wave One (Rainier) was the first quantum annealing (a.k.a. adiabatic quantum optimization) processor implementing the Ising model using 128 superconducting quantum bits (qubits). With the CO problem mapped onto qubits, 
the quantum annealer is claimed to benefit from quantum tunneling and fluctuations to find its ground state.

Despite impressive efforts, all demonstrated IM implementations face serious shortcomings.
Rainier and all its successors, such as the latest D-Wave 2000Q (with 2048 qubits), operate at millikelvin temperature, requiring large cryogenic facilities with kWs of power consumption. The technology has also not scaled well to larger IMs, comes with a very high production cost, and suffers from a relatively sparse qubit connectivity (six).

Coherent Ising Machines (CIM) implemented using externally pumped optical parametric oscillators (OPO)\cite{inagaki2016sci,marandi2014natp} are room-temperature alternatives, which both avoids the cryogenic requirements and considerably outperform D-Wave's most advanced machine 2000Q in solving large scale Max-Cut problems\cite{hamerly2019sciadv}. However, they too face serious challenges in terms of scaling, footprint, and high power consumption. 
In addition, CIMs require optical table infrastructure and kilometer-long optical fibers in order to accommodate all time-multiplexed optical parametric oscillators. Once, these systems are built and the operating frequency is set, no further tuning is available unless new hardware with a new architecture replaces the existing one.

As a room-temperature digital CMOS alternative, Fujitsu, in 2018, announced its state-of-the-art field-programmable digital array (FPGA) digital annealer unit (DAU), offering 8192 simulated qubits with all--to--all connectivity allowing all qubits to exchange signals freely. Thanks to their superior connectivity, DAUs are capable of dealing with sizeable real-world scale CO problems. However, they are still relying on the Von-Neumann computing paradigm with the added reconfigurability of FPGAs and not the nature of a physical system. It is hence not likely that this approach will be able to scale to much larger systems.

Inspired by both the optical pumping mechanism, and the practical nature of ordinary CMOS electronics, 
Tianshi Wang and Jaijeet Roychowdhury\cite{wang2019oim} proposed non-linear oscillator--based IMs based on off-the-shelf electronics. As they have demosntrated, the phase dynamics of a network of coupled oscillators can minimize a scalar function called a Lyapunov function, which serves as a measure of the network stability. Once the networks is under second-harmonic injection locking (SHIL), the individual phase of any oscillators in the network can only obtain binarized values, \emph{i.e.,} 0 or $\pi$. Under SHIL, the Lyapunov function directly translates into the Ising Hamiltonian, and the network can operate as an IM. Annealing can be implemented in different ways, \emph{e.g.} through varying the strength of the injected signal. The authors demonstrated a PCB prototype using a network of 240 resistively coupled CMOS LC oscillators\cite{wang2019oim1} with a maximum of 1200 connections, achieving a remarkable solution time of 1~ms for a moderate operating frequency of 1 MHz and a very modest power consumption of only 5~W. 
In a similar demonstration a network of four all-to-all LC coupled oscillators was shown to solve Max-Cut problems of small size~\citen{chou2019scirep}. Both reports forecast a significant boost in processing speed and solutions quality if the CMOS coupled oscillators could be realized in large numbers and operating at GHz frequencies. . 
One of the most intriguing properties of SHNOs \cite{liu2012prl,liu2013prl,demidov2012nm,Demidov2014apl,ranjbar2014ieeeml, Zahedinejad2018apl,Mazraati2016apl,chen2016ieee} 
is their ability to not only mutually synchronize in both one\cite{awad2017ntphys} and two\cite{Zahedinejad2020natnano} dimensions, but also to an external source \cite{demidov2014ntc,Hache2019APL}, clearly demonstrating strong inter-oscillator coupling \cite{Dvornik2018prappl,Spicer2018prb,ulrichs2014apl}. Other related synchronized nano-oscillators, such as spin torque nano-oscillators (STNOs) were recently used for vowel recognition at about 300 MHz.\cite{romera2018nature} However, scaling up to much larger STNO arrays have proven difficult due to the slow progress in number of mutually synchronized STNOs.\cite{mancoff2005nt,kaka2005nt,sani2013ntc,Houshang2015natnano} As SHNOs have been demonstrated down to lateral dimensions of only 20 nm\cite{durrenfeld2017nanoscale}, demonstrate flexibility and wide-range tunability\cite{Fulara2019SciAdv,Awad2020arxiv,mazraati2018apl,divinskiy2017apl,Zahedinejad2017ieeeml}, operated up to 26 GHz\cite{Zahedinejad2018apl, Fulara2019SciAdv}, showing mutual synchronization in two-dimensional arrays of up to 64 oscillators\cite{Zahedinejad2020natnano}, they hence represent the most promising route towards truly miniaturized, ultra-fast, and large-scale oscillator based Ising Machines.

Here we present the world's most miniaturized, and the first SHNO based, Ising machine. We demonstrate robust phase binarization of both 1 $\times$ 2 and 2 $\times$ 2 SHNO arrays using second-harmonic microwave current injection locking. The phase binarization manifests itself as distinct microwave output power levels, which are readily distinguised using electrical means. In addition, we use phase-resolved Brillouin Light Scattering (phase-BLS) microscopy to directly observe the individual phases of the precessing magnetization in each nano-constriction. The different high/low microwave output states can be directly mapped onto different in-
and anti-phase states in both types of array, and, as expected, an additional intermediate power mixed-phase state in the 2 $\times$ 2 array. The different states can be accessed using either different injected power levels or a detuned frequency of the injected signal.  

\section*{Results}

\begin{figure}
  \begin{center}
  \includegraphics[width=8cm]{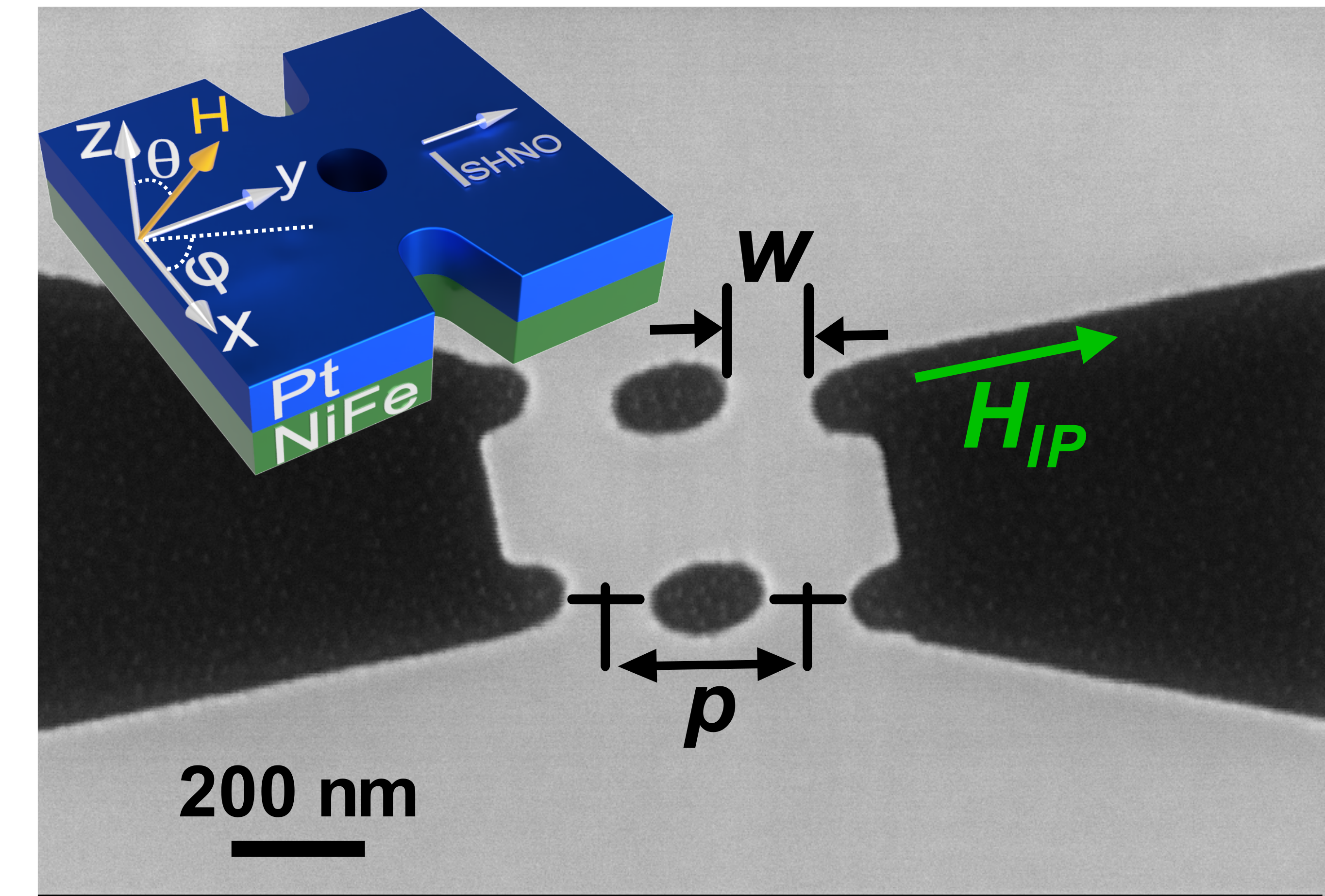}
    \caption{A scanning electron microscopy of a 2x2 SHNO array is shown together with a schematic for a 1x2 array. SHNOs are made of a Pt/NiFe bilayer, 5 and 3 nm thick, respectively. The Hf dusting layer is omitted here. Width of each individual SHNO is W = 120 nm and the pitch size is P = 300 nm. $H_{IP}$ shows the in-plane direction of the external magetic field, H.}
    \label{fig:1}
    \end{center}
\end{figure}

A schematic illustration of a 1x2 SHNO array is shown in Fig.1 together with an SEM image of a 2x2 array. All devices are made of a bilayer of Pt(5)/NiFe(3) (numbers in parentheses indicated the thickness in nm) in which dusting the interface with an ultra thin Hf layer of 0.5 nm reduces the damping in NiFe \cite{mazraati2018apl}. The width of all nano-constrictions is $w=$ 120 nm and the pitch (separation) is $p=$ 200 nm for the 1x2 array and $p=$ 300 nm for the 2x2 array (see the detailed fabrication method in \citen{Zahedinejad2020natnano}). A magnetic field $H$ is applied with an out-of-plane angle $\theta$ and an in-plane angle $\phi$; throughout this study $\theta=$ 78$^\circ$ and $\phi=$ 24$^\circ$. In the SEM image the in-plane component $\mathit{H_{IP}}$ is explicitely shown. 
$\mathit{I_{SHNO}}$ shows the direction of the charge current. 
  
\begin{figure}
  \begin{center}
  \includegraphics[width=0.8\textwidth]{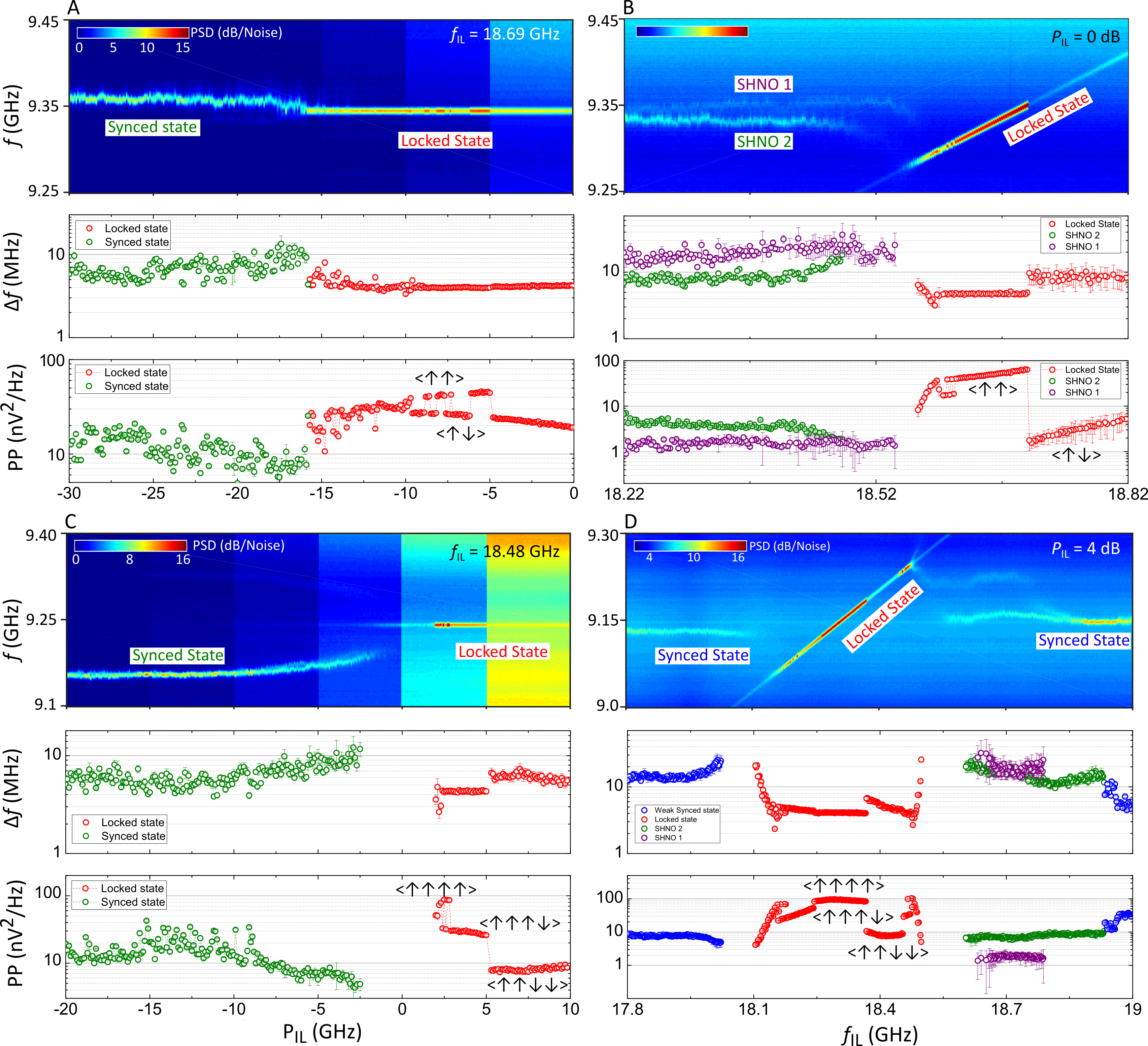}
    \caption{PSD of a 1x2 SHNO array (a) as a function of injection power, $P_{IL}$ at a fixed injected frequency $f_{IL}$ = 18.69 GHz, and (b)  as a function of injection frequency at a fixed $P_{IL}$= 0 dB. In (a) the array starts off in a syncronized state (Synced state) while in (b) the SHNOs are not synchronized. In both cases, after the array is injection locked to the external source, the linewidth decreases and the power shows intermittent fluctuations between a high and a low power state. The same trend is observed in 2x2 arrays. However, in the 2x2 arrays, three different energy levels can be distinguished once the array is locked to the external source as shown in (c) and (d).}
    \label{fig:2}
    \end{center}
\end{figure}

Fig 2.(a) shows the power spectral density (PSD) of the 1x2 array, running at $\mathit{I_{SHNO}}$ = 3 mA in a field of $H$ = 6400 Oe, as the injected power, $\mathit{P_{IL}}$, is ramped from --30 to 0 dBm. The SHNOs start out in their mutually synchronized state. When $\mathit{P_{IL}}$ reaches a threshold of about --16 dBm 
the SHNOs lock intermittently to the external source. This injection locking grows gradually stronger and is relatively stable above --13 dBm, as evidenced by a substantially lower and stable linewidth and much higher and stable peak power. 
When the injected power is further increased to --10 dBm, the peak power splits into a high and a low power branch without any significant change in the linewidth. The two branches increasingly diverge with increasing $\mathit{P_{IL}}$ and the peak power switches a few times between its two states on a slow timescale while remaining stable within each branch in between jumps. The timescale for the switching appears to be slowing down with increasing $\mathit{P_{IL}}$ and above --5 dBm the SHNO remains in the lower branch for the remainder of the measurement. 

The observed behavior is very close to that predicted for phase binarization in oscillator networks under second-harmonic injection locking (SHIL)\cite{wang2019oim}. 
For an individual oscillator, there exist two degenerate injection locked states. For two or more oscillators with either inter-oscillator coupling or slightly different intrinsic frequencies (or both), this degeneracy can be lifted as it might be energetically favorable for the two oscillators to injection lock out-of-phase. With increasing $\mathit{P_{IL}}$, the energy difference between these two states increases, in agreement with the slower hopping and final stability observed in Fig.2A. With increasing $\mathit{P_{IL}}$ the preferred relative phase between the two oscillators in their mutually synchronized state should also be overcome by the injected signal, such that the in-phase locked state, which we from hereon denote as $<\uparrow\uparrow>$, increases in power (the relative phase decreases), and the out-of-phase state, $<\uparrow\downarrow>$, decreases in power (further increasing relative phase). This is in direct agreement with the observed diverging nature of the two branches.

Fixing the injected power and instead sweeping the injected frequency reproduces the same phase binarization phenomenon in Fig.2B. In this measurement, the two SHNOs are not mutually synchronized and hence run on slightly different intrinsic frequencies. 
However, as the injected signal approaches the frequency of the SHNOs, initially one and then both of the oscillators get locked and after going through a region of unstable fluctuations, the SHNOs first fall into the $<\uparrow\uparrow>$ state and eventually switches into the $<\uparrow\downarrow>$ state. It is noteworthy that the system switches from $<\uparrow\uparrow>$ to $<\uparrow\downarrow>$ when the injected frequency increases beyond the intrinsic frequencies of both oscillators instead of being in between them. In addition, the difference in peak power between the two states is now close to two orders of magnitude, indicating a close to perfect cancellation of the microwave signal in the $<\uparrow\downarrow>$ state.

We now turn to the two-dimensional 2x2 arrays where similar phase binarization can be observed with yet more phase binarized states appearing. As seen in Fig.2C, a similar trend can be observed as for the 1x2 array, \emph{i.e.}~the four SHNOs start out in a mutual synchronized state, which injection locks to the external signal at about --2 dBm. Here, the microwave signal all but disappears, which we interpret as a complete anti-phase state of the type $<\uparrow\uparrow\downarrow\downarrow>$, or any of its equivalent degenerate states. In a tiny region around 2 dBm the array switches into a high power state, which we interpret as $<\uparrow\uparrow\uparrow\uparrow>$. However, this high power state is soon replaced by an intermediate state with about 1/4 of the power, consistent with a $<\uparrow\uparrow\uparrow\downarrow>$ state or any of its degenerate states. At 5 dBm a very low-power anti-phase state is again reached. Just as for the the 1x2 array, the anti-phase state is favored at high injectded power. 

\begin{figure}
  \begin{center}
  \includegraphics[width=0.45\textwidth]{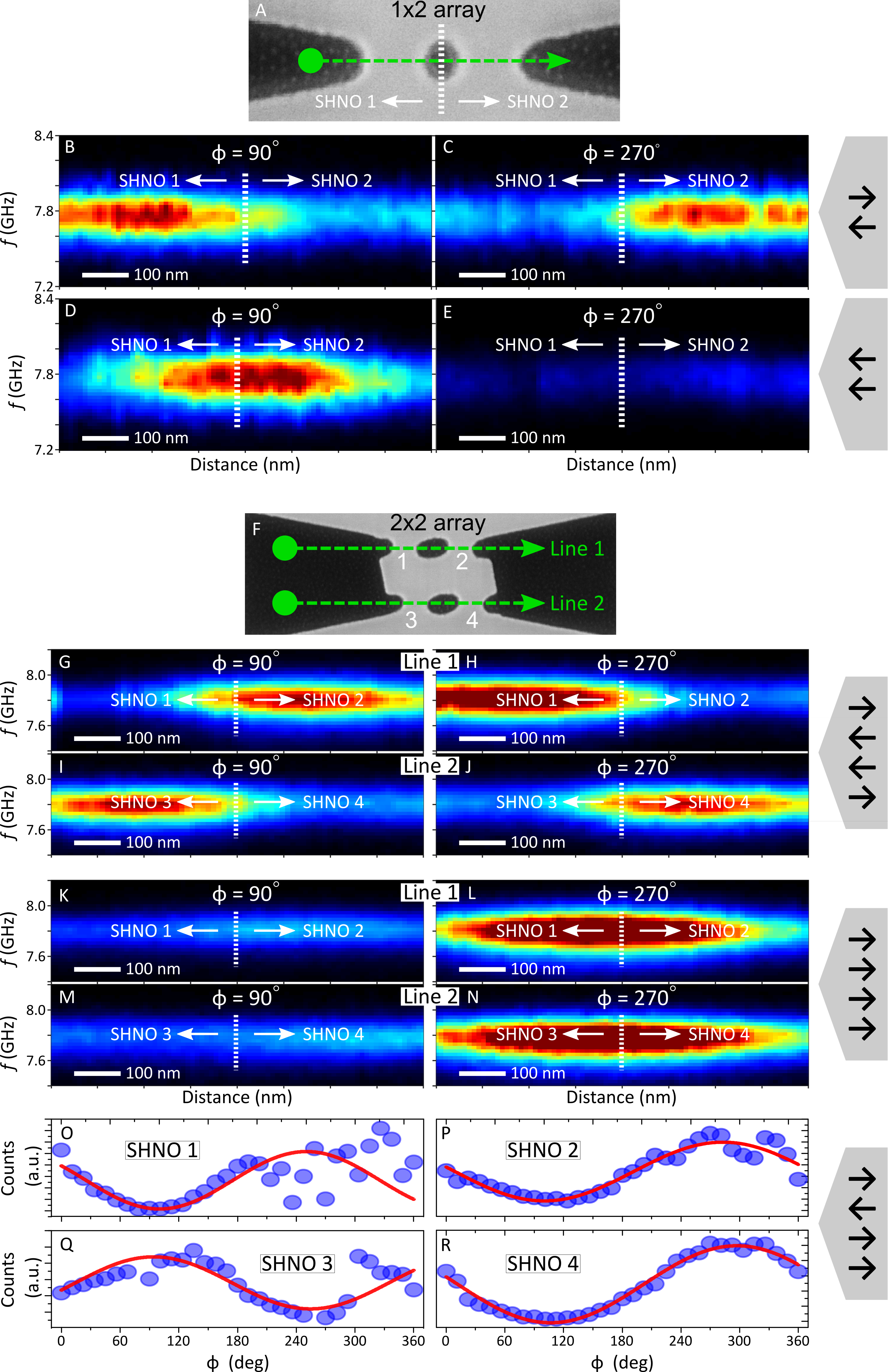}
    \caption{
    (\textbf{A}) SEM image of 1$\times$2 SHNO array 
    while under SHIL with $P_{\mathrm{IL}}$ = 0 dBm and $f_{\mathrm{IL}}$ = 15.6 GHz. The BLS scan direction is shown by the green dashed arrow. (\textbf{B})-(\textbf{E})  show the SW intensity profile of the SHNOs for phase--resolved BLS signal at 90$^\circ$ and 270$^\circ$. 
    SHNOs in (\textbf{B}) and (\textbf{C}) are energized with 180$^\circ$ phase difference ($< \uparrow \downarrow >$) when $2f_0$ $<$ $f_{\mathrm{IL}}$ and in (\textbf{D}) and (\textbf{E}) with 0$^\circ$ phase difference ($< \uparrow \uparrow >$) when $2f_0$ $\simeq$ $f_{\mathrm{IL}}$. (\textbf{F}) SEM image of a 2$\times$2 SHNO array with corresponding BLS line scan directions. 
    Phase-resolved BLS line scans are shown in (\textbf{G})-(\textbf{J}) for $P_{\mathrm{IL}}$ = +4 dBm corresponding to $< \downarrow \uparrow \uparrow \downarrow >$ phase state while (\textbf{K})-(\textbf{N}) shows $< \downarrow \downarrow \downarrow \downarrow >$ state obtained at $P_{\mathrm{IL}}$ = +3 dBm. (\textbf{O})-(\textbf{R}) BLS counts for the least stable phase state $<\downarrow \downarrow \uparrow \downarrow>$ at $P_{\mathrm{IL}}$ = -2 dBm measured by sweeping the phase ($\Phi$) of the BLS.}
    \label{fig:3}
    \end{center}
\end{figure}

To gain further insight into these different states and allow us to determine the exact phase of each nano-constriction inside the arrays, we now turn to phase-sensitive Brillouin Light Scattering (phase-BLS) microscopy (see Methods for details). Fig.\ref{fig:3}A-E show the phase-BLS results from the 1$\times$2 SHNO array biased at $I_{\mathrm{dc}}$ = 2.8 mA in a magnetic field $H$ = 5930 Oe, resulting in a free-running frequency of $f_0 \simeq$ 7.8 GHz. 
The array is under SHIL with $P_{\mathrm{IL}}$ = 0 dBm and $f_{\mathrm{IL}}$ = 15.6 GHz. Fig.\ref{fig:3}A shows an SEM image of the array with dashed green arrows indicating the direction of a phase-BLS line scan across both nano-constrictions. 
Fig.\ref{fig:3}B shows the phase-BLS counts vs.~location for the electrically determined $<\uparrow\downarrow>$ state, where the phase-BLS detection angle is set to 90$^\circ$: The line scan shows a high SW intensity only in SHNO 1. When the phase-BLS detection angle is rotated 180$^\circ$ as in Fig.\ref{fig:3}B, the counts are instead located in SHNO 2. The electrically measured $<\uparrow\downarrow>$ state can hence be directly mapped onto an anti-phase state in the auto-oscillations of SHNO 1 and 2. Similarly, when we slightly change the experimental conditions to realize an electrical $<\uparrow\uparrow>$ state, the same type of phase-BLS line scans now clearly show in Fig.\ref{fig:3}D and E that both SHNOs are exactly in-phase with each other.

Similar line scans can now be made for the 2$\times$2 SHNO arrays as shown in Fig.\ref{fig:3}F. In Fig.\ref{fig:3}G-J, the array was supplied with $I_{\mathrm{dc}}$ = 3 mA in a field $H$ = 5980 Oe, and $P_{\mathrm{IL}}$ = +4 dBm was used for SHIL; the detected electrical microwave signal was vanishingly small, indicating an anti-phase state of the array. 
The two phase-BLS line scans directly corroborates this picture as the SHNOs 1 and 4 have 0$^\circ$ relative phase difference and operate at 180$^\circ$ phase difference with SHNOs 2 and 3. This  arrangement corresponds to a $<\downarrow\uparrow\uparrow\downarrow>$ phase state which is reflected as the low peak power state in the electrical measurement. When we increase $H$ to 6000 Oe, we find from the electrical measurements that we favor a fully in-phase $<\downarrow\downarrow\downarrow\downarrow>$ state. This is again directly corroborated by the phase-BLS scans in Fig.\ref{fig:3}K-N. 

In addition, we can also realize the intermediate electrical state and now directly observe which SHNO is out-of-phase with the others.  
Instead of showing line scans, we here want to highlight the sinusoidal phase dependence of the phase-BLS counts as a function of relative injection locking phase angle. We have measured the phase-BLS counts in four spatial locations corresponding to each nano-constriction center. Varying the phase angle from 0 to 360$^\circ$ we clearly observe that SHNO 1, 2, and 4 show the same type of sinusoidal dependence of BLS counts vs.~phase angle, whereas SHNO 3 shows exactly the opposite behavior. In other words, phase-BLS allows us to determine the exact $<\downarrow\downarrow\uparrow\downarrow>$ state out of the four degenerate possibilities for the intermediate state.   

As SHNO arrays as large as 10x10 have been shown to be fully operational with strong inter-oscillator coupling, we argue that our present demnostration should be possible to scale up to very large arrays. We also believe that voltage-controlled magnetic anisotropy in CoFeB based SHNOs\cite{Fulara2019SciAdv} should be a possible route towards a fully integrated voltage-controlled SHNO Ising Machine. For fast electrical read-out, one should also add magnetic tunnel junctions to the auto-oscillating regions of each nano-constriction. While these are important challenges, the very high operating frequency, the very high degree of miniaturization, the demonstrated scaling path, and the CMOS compatibility, all hold tremendous promise for a ground-breaking technology platform towards wide-purpose Ising Machines.

\begin{methods}

\subsection{Microwave measurements and SHIL.}
Microwave measurements were carried out at room temperature using a custom-built probe station where the sample was mounted at a fixed IP and OOP angle of $\phi$ = 24$^{\circ}$ and $\theta$ = 76$^{\circ}$. A direct positive electric current, $I_\text{dc}$, was then injected through the \textit{dc} port of a high-frequency bias-T to induce auto-oscillations. Through microwave circulator that was also connected to the high frequency port of the SA, a RF current from a Rhode \& Schwarz (10 Hz-20 GHz) signal source, was injected into the auto-oscillating sample. The resulting signal was then amplified by a low-noise amplifier with a gain of $\geq$ 56 dB (4-10 GHz) and subsequently recorded using a spectrum analyzer (SA) from Rhode \& Schwarz (10 Hz-40 GHz) comprising a low-resolution bandwidth of 300 kHz. A schematic of the measurement setup is shown in the work of Zahedinejad \textit{et.al.} \cite{Zahedinejad2017ieeeml}.

\subsection{Phase--resolved BLS microscopy.}
 Phase--resolved BLS microscopy was performed by modulating the phase of the incoming light using an eletro-optic modulator (EOM) at half the frequency (f) as that of the injection signal (2f) to the SHNO array. The inelastically scattered outgoing light, carrying the phase information from the oscillator, interferes with the elastically scattered light beam whose phase ($\phi$) can be controlled by a phase shifter (resolution of 5.6$^{\circ}$) attached to the EOM. The resulting BLS signal will then be the difference in phase of the light (known) and the oscillator (unknown). The signal was collected and analyzed in a 6-pass Tandem Fabry-Perot Interferometer (TFPI) and detected using a single channel avalanche photodiode. High spatial resolution is obtained at the expense of wavevector resolution by using a high N.A. microscope objective which forms a tight focus of the probe beam down to the diffraction limit (beam diameter $\approx$ 300~nm). A 3-axis nanometer-resolution stage along with active stabilization using a software provided by THATec Innovation GmbH has made it possible to perform very precise linear scans.

\end{methods}

\begin{addendum}
 \item This work was partially supported by the Horizon 2020 research and innovation programme (ERC Advanced Grant No.~835068 "TOPSPIN"). This work was also partially supported by the Swedish Research Council (VR) and the Knut and Alice Wallenberg Foundation. J.Ch. acknowledges funding from program No. 2017/24/T/ST3/00009 by National Science Center, Poland. Numerical calculations were supported in part by the PL-Grid infrastructure.
  \item[Author contributions] M.Z. designed and fabricated the devices. A.H. carried out all the electrical measurements and analyzed them together with M.Z. S.M. and A.A.A. carried out all optical measurements and their analysis. 
  J.\AA. initiated and supervised the project. All authors contributed to the analysis and interpretation of the results and co-wrote the manuscript.
 \item[Correspondence] Correspondence and requests for materials
should be addressed to J. \AA kerman~(e-mail: johan.akerman@physics.gu.se).
\end{addendum}

\end{document}